\PassOptionsToPackage{unicode}{hyperref}
\PassOptionsToPackage{hyphens}{url}
\PassOptionsToPackage{dvipsnames,svgnames,x11names}{xcolor}
\documentclass[
]{article}
\usepackage{amsmath,amssymb}
\usepackage{lmodern}
\usepackage{iftex}
\ifPDFTeX
  \usepackage[T1]{fontenc}
  \usepackage[utf8]{inputenc}
  \usepackage{textcomp} 
\else 
  \usepackage{unicode-math}
  \defaultfontfeatures{Scale=MatchLowercase}
  \defaultfontfeatures[\rmfamily]{Ligatures=TeX,Scale=1}
\fi
\IfFileExists{upquote.sty}{\usepackage{upquote}}{}
\IfFileExists{microtype.sty}{
  \usepackage[]{microtype}
  \UseMicrotypeSet[protrusion]{basicmath} 
}{}
\makeatletter
\@ifundefined{KOMAClassName}{
  \IfFileExists{parskip.sty}{%
    \usepackage{parskip}
  }{
    \setlength{\parindent}{0pt}
    \setlength{\parskip}{6pt plus 2pt minus 1pt}}
}{
  \KOMAoptions{parskip=half}}
\makeatother
\usepackage{xcolor}
\usepackage{graphicx}
\makeatletter
\def\maxwidth{\ifdim\Gin@nat@width>\linewidth\linewidth\else\Gin@nat@width\fi}
\def\maxheight{\ifdim\Gin@nat@height>\textheight\textheight\else\Gin@nat@height\fi}
\makeatother
\setkeys{Gin}{width=\maxwidth,height=\maxheight,keepaspectratio}
\makeatletter
\def\fps@figure{htbp}
\makeatother
\providecommand{\tightlist}{%
  \setlength{\itemsep}{0pt}\setlength{\parskip}{0pt}}
\NewDocumentCommand\citeproctext{}{}
\NewDocumentCommand\citeproc{mm}{%
  \begingroup\def\citeproctext{#2}\cite{#1}\endgroup}
\makeatletter
 \let\@cite@ofmt\@firstofone
 \def\@biblabel#1{}
 \def\@cite#1#2{{#1\if@tempswa , #2\fi}}
\makeatother
\newlength{\cslhangindent}
\newlength{\csllabelwidth}
\newenvironment{CSLReferences}[2] 
 {\begin{list}{}{%
  \setlength{\itemindent}{0pt}
  \setlength{\leftmargin}{0pt}
  \setlength{\parsep}{0pt}
  \ifodd #1
   \setlength{\leftmargin}{\cslhangindent}
   \setlength{\itemindent}{-1\cslhangindent}
  \fi
  \setlength{\itemsep}{#2\baselineskip}}}
 {\end{list}}
\usepackage{calc}

\ifLuaTeX
\usepackage[bidi=basic]{babel}
\else
\usepackage[bidi=default]{babel}
\fi
\babelprovide[main,import]{american}

\def\languageshorthands#1{}
\ifLuaTeX
  \usepackage{selnolig}  
\fi
\IfFileExists{bookmark.sty}{\usepackage{bookmark}}{\usepackage{hyperref}}
\IfFileExists{xurl.sty}{\usepackage{xurl}}{} 
\hypersetup{
  pdftitle={zea: A Toolbox for Cognitive Ultrasound Imaging},
  pdfauthor={Tristan S. W. Stevens, Wessel L. van Nierop, Ben Luijten,
Vincent van de Schaft, Oisín Nolan, Beatrice Federici, Louis D. van
Harten, Simon W. Penninga, Noortje I. P. Schueler, Ruud J. G. van
Sloun},
  pdflang={en-US},
  colorlinks=true,
  linkcolor={Maroon},
  filecolor={Maroon},
  citecolor={Blue},
  urlcolor={Blue},
  pdfcreator={LaTeX via pandoc}}

\title{zea: A Toolbox for Cognitive Ultrasound Imaging}

\definecolor{c53baa1}{RGB}{83,186,161}
\definecolor{c202826}{RGB}{32,40,38}

\usepackage[affil-it]{authblk}
\usepackage{orcidlink}
\author[1%
  \ensuremath\mathparagraph]{Tristan S. W. Stevens%
    \,\orcidlink{0000-0002-8563-5931}\,%
    }
\author[1%
  ]{Wessel L. van Nierop%
    \,\orcidlink{0009-0003-3141-3369}\,%
    }
\author[1%
  ]{Ben Luijten%
    \,\orcidlink{0000-0002-1797-8721}\,%
    }
\author[1%
  ]{Vincent van de Schaft%
    \,\orcidlink{0000-0002-8515-5372}\,%
    }
\author[1%
  ]{Oisín Nolan%
    \,\orcidlink{0009-0002-6939-7627}\,%
    }
\author[1%
  ]{Beatrice Federici%
    \,\orcidlink{0009-0003-2496-8825}\,%
    }
\author[1%
  ]{Louis D. van Harten%
    \,\orcidlink{0000-0002-0943-2825}\,%
    }
\author[1%
  ]{Simon W. Penninga%
    \,\orcidlink{0009-0003-4095-8168}\,%
    }
\author[1%
  ]{Noortje I. P. Schueler%
    \,\orcidlink{0009-0003-7134-6850}\,%
    }
\author[1%
  ]{Ruud J. G. van Sloun%
    \,\orcidlink{0000-0003-2845-0495}\,%
    }

\affil[1]{Eindhoven University of Technology, the Netherlands%
  }
\affil[$\mathparagraph$]{Corresponding author: %
}
\date{20 June 2025}

\begin{document}
\maketitle

\section{Summary}\label{summary}

\begin{figure}
\centering
\includegraphics[width=1\linewidth,height=\textheight,keepaspectratio]{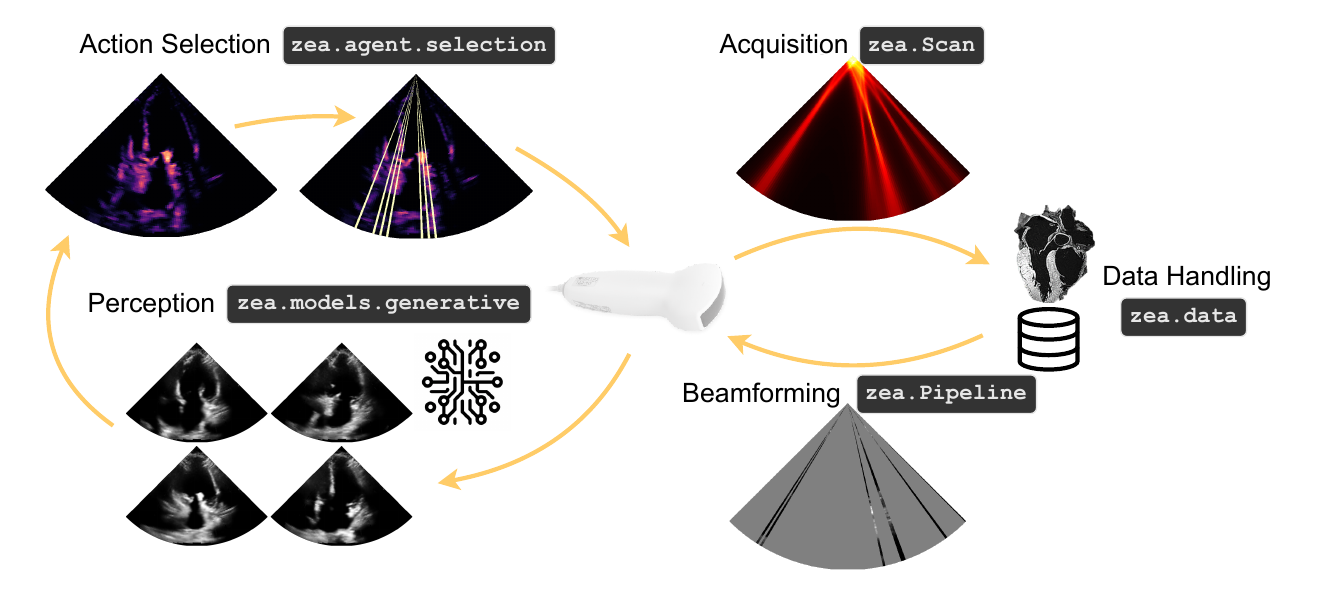}
\caption{High-level overview of an ultrasound perception-action loop
implemented in zea.\label{fig:diagram}}
\end{figure}

Ultrasound imaging is a powerful medical imaging modality that is widely
used in clinical settings for various applications, including
obstetrics, cardiology, and abdominal imaging. While ultrasound imaging
is non-invasive, real-time, and relatively low-cost compared to other
imaging modalities such as MRI or CT, it still faces challenges in terms
of image quality and interpretation. Many signal processing steps are
required to extract useful information from the raw ultrasound data,
such as filtering, beamforming, and image reconstruction. Traditional
ultrasound imaging techniques often suffer from reduced image quality as
naive assumptions are made in these processing steps, which do not
account for the complex nature of ultrasound signals. Furthermore,
acquisition (action) and reconstruction (perception) of ultrasound is
often performed disjointly. Cognitive ultrasound imaging
(\citeproc{ref-van2024active}{Sloun, 2024}), see \autoref{fig:diagram},
is a novel approach that aims to address these challenges by leveraging
more powerful generative models, enabled by advances in deep learning,
to close the action-perception loop. This approach requires a redesign
of current common ultrasound imaging pipelines, where parameters are
expected to be changed dynamically based on past and current
observations. Furthermore, the high-dimensional nature of ultrasound
data requires powerful deep generative models to learn the structured
distribution of ultrasound signals and to effectively solve inverse
problems that capture the challenges of ultrasound imaging
(\citeproc{ref-stevens2025deep}{Stevens, Overdevest, et al., 2025}).
This necessitates a flexible and efficient toolbox that can handle the
complexities of cognitive ultrasound imaging, including a real-time
ultrasound reconstruction pipeline, dynamic parameter adjustment, and
advanced generative modeling.

We present \texttt{zea} (pronounced \emph{ze-yah}), a Python package for
cognitive ultrasound imaging that provides a flexible, modular and
differentiable pipeline for ultrasound data processing, as well as a
collection of pre-defined models for ultrasound image and signal
processing. The toolbox is designed to be easy to use, with a high-level
interface that allows users to define their own ultrasound
reconstruction pipelines, and to integrate deep learning models into the
pipeline. The toolbox is built on top of Keras 3
(\citeproc{ref-chollet2015keras}{Chollet \& others, 2015}), which
provides a framework for building and training deep learning models with
the three major deep learning frameworks as backend: TensorFlow
(\citeproc{ref-abadi2016tensorflow}{Abadi et al., 2016}), PyTorch
(\citeproc{ref-NEURIPS2019_9015}{Paszke et al., 2019}) and JAX
(\citeproc{ref-jax2018github}{Bradbury et al., 2018}). This means that
it is easy to integrate a custom ultrasound reconstruction pipeline in a
machine learning workflow. In the past few years, several works have
used and contributed to \texttt{zea}, including Luijten et al.
(\citeproc{ref-luijten2020adaptive}{2020}), Schaft et al.
(\citeproc{ref-van2024off}{2025}), Stevens et al.
(\citeproc{ref-stevens2024dehazing}{2024}), Nolan et al.
(\citeproc{ref-nolan2024active}{2025}), Federici et al.
(\citeproc{ref-federici2024active}{2025}), Stevens, Nolan, Robert, et
al. (\citeproc{ref-stevens2025sequential}{2025}), Penninga et al.
(\citeproc{ref-penninga2025deep}{2025}) and Stevens, Nolan, Somphone, et
al. (\citeproc{ref-stevens2025high}{2025}).

\section{Statement of need}\label{statement-of-need}

The ultrasound research community has advanced significantly due to
publicly available high-quality software, including simulation tools
such as \texttt{Field\ II} (\citeproc{ref-jensen2004simulation}{Jensen,
2004}) and \texttt{k-wave} (\citeproc{ref-treeby2010k}{Treeby \& Cox,
2010}), as well as reconstruction and real-time processing libraries
like \texttt{USTB}
(\citeproc{ref-rodriguez2017ultrasound}{Rodriguez-Molares et al.,
2017}), \texttt{MUST} (\citeproc{ref-garcia2021make}{Garcia, 2021}),
\texttt{ARRUS} (\citeproc{ref-jarosik2020arrus}{Jarosik \& others,
2020}), \texttt{FAST} (\citeproc{ref-smistad2021fast}{Smistad, 2021}),
\texttt{QUPS} (\citeproc{ref-brevett2024qups}{Brevett, 2024}), and
\texttt{vbeam} (\citeproc{ref-magnus2023vbeam}{Kvalevåg et al., 2023}).
However, existing solutions are not well-equipped for cognitive
ultrasound imaging, where the integration of deep learning and dynamic,
closed-loop ultrasound reconstruction pipelines is essential. Our aim
with \texttt{zea} is to provide a complementary, highly flexible and
differentiable pipeline written in a modern deep learning framework, as
well as offer a convenient platform for pretrained models. This
addresses the need for a modular and extensible library that supports
cognitive ultrasound workflows and seamless integration with
state-of-the-art machine learning models. While the full realization of
cognitive ultrasound imaging remains an ongoing effort, we hope this
toolbox will help spur further research and development in the field.

\section{Overview of functionality}\label{overview-of-functionality}

\texttt{zea} is an open-source Python package, available at
\url{http://github.com/tue-bmd/zea}, that consists of the following core
components:

\begin{itemize}
\tightlist
\item
  \textbf{Data}: A set of data handling classes such as
  \texttt{zea.File}, \texttt{zea.Dataset} and \texttt{zea.Dataloader},
  suited for machine learning workflows. \texttt{zea} works with HDF5
  files, storing data and acquisition parameters together in a single
  file, which can be easily loaded and saved through the \texttt{zea}
  API. For more demanding workflows, such as training deep learning
  models, \texttt{zea} offers robust data loading utilities such as
  batching, shuffling, caching, and preprocessing. Additionally, we
  provide examples and conversion scripts for popular ultrasound
  datasets, such as CAMUS (\citeproc{ref-leclerc2019deep}{Leclerc et
  al., 2019}), PICMUS (\citeproc{ref-liebgott2016plane}{Liebgott et al.,
  2016}), and EchoNet (\citeproc{ref-ouyang2020video}{Ouyang et al.,
  2020}).
\item
  \textbf{Pipeline}: A modular and differentiable pipeline class that
  allows users to define a sequence of operations
  (\texttt{zea.Operation}) to process ultrasound data. The pipeline is
  stateless and supports \emph{Just in Time} (JIT) compilation.
  Ultimately, this allows for dynamic parameter adjustment, as well as
  real-time integration of deep learning models inside the ultrasound
  reconstruction pipeline.
\item
  \textbf{Models}: A collection of pre-defined models for ultrasound
  image and signal processing. Similar to the data, these models can be
  loaded locally or from the
  \href{https://huggingface.co/zeahub}{Hugging Face Hub}. Besides
  supervised models, \texttt{zea} also provides a set of (deep)
  generative models, with an interface to solve inverse problems in
  ultrasound imaging within a probabilistic machine learning framework.
\item
  \textbf{Agents}: A set of tools to interact with the pipeline and
  models. These agents can be used to alter the pipeline parameters, or
  select a subset of acquired data. The agent module closes the
  action-perception loop (\citeproc{ref-van2024active}{Sloun, 2024}),
  tying together acquisition and reconstruction of ultrasound data.
\end{itemize}

For detailed examples and use cases, please refer to the example
notebooks available on the documentation:
\url{https://zea.readthedocs.io/}.

\section{Availability, Development, and
Documentation}\label{availability-development-and-documentation}

\texttt{zea} is available through PyPI via \texttt{pip\ install\ zea},
and the development version is available via GitHub. GitHub Actions
manage continuous integration through automated code testing (PyTest),
code linting and formatting (Ruff), and documentation generation
(Sphinx). The documentation is hosted on ReadTheDocs. At the time of
writing, 20 example notebooks are available, covering the various
discussed components of the toolbox. The package is licensed under the
Apache License 2.0, which allows for both academic and commercial use.

\section*{References}\label{references}
\addcontentsline{toc}{section}{References}

\phantomsection\label{refs}

\end{document}